\documentclass[twocolumn,prl,showpacs]{revtex4}

\usepackage{graphicx}
\usepackage{rotating}
\usepackage{amsmath}
\usepackage{amsfonts}
\usepackage{amssymb}
\usepackage{enumerate}
\usepackage{longtable}
\usepackage{dcolumn}
\usepackage{bm}

\begin{document}

\newcommand{\be}{\begin{equation}}
\newcommand{\ee}{\end{equation}}
\newcommand{\bn}{\begin{eqnarray}}
\newcommand{\en}{\end{eqnarray}}

\title{Mott Transition and Strange Metal in Two Dimensions: 
A View from Cellular Dynamical Cluster Approximation}

\author{M. S. Laad}
\email{mukul@mpipks-dresden.mpg.de}
\affiliation{Max-Planck-Institut f\"ur Physik Komplexer Systeme, 38 N\"othnitzer Strasse, 
 01187 Dresden, Germany }
\author{L. Craco}
\affiliation{Institut f\"ur Chemische Physik Fester Stoffe, 
40 N\"othnitzer Strasse, 01187 Dresden, Germany}
  
\date{\rm\today}

\begin{abstract}
We introduce a Cellular Dynamical Cluster Approximation (CDCA) to 
study the nature of the Mott insulator-metal transition in the extended 
Hubbard model on a square lattice.  At strong coupling, a $d$-wave Mott 
insulator is obtained. Hole doping drives a first order Mott transition 
to a non-Fermi (nFL) liquid metal.  Remarkably, this nFL is caused by an 
Anderson orthogonality catastrophe at low energies due to the non-trivial 
competition between strong, non-local interactions and hopping. This 
constitutes the first explicit realisation of Anderson's Luttinger liquid 
idea in two dimensions. Many experimental responses in the ``strange metal''
phase found around optimal doping in cuprates
are understood naturally within our approach.
\end{abstract}
     
\pacs{71.30.+h,72.10.-d,74.20.Mn}

\maketitle

The problem of high-$T_{c}$ superconductivity (HTSC) in quasi-two-dimensional
cuprates remains an outstanding problem of modern condensed matter physics.
HTSC arises from a new, non-Fermi liquid (nFL) state, characterised by 
highly anomalous normal state response of the {\it underdoped} cuprates, 
continuing up to optimal doping, $x_{opt}$ (where the SC $T_{c}$ is maximum).  
Beyond $x_{opt}$, a low-$T$ Fermi liquid state seems to be 
recovered~\cite{[1]}. $(i)$ Superconductivity itself is unconventional, 
and of $d$-wave symmetry~\cite{[2]}.  Subsequently, it was also found that 
the normal state pseudogap (in the one-particle spectral function) possesses 
$d$-wave symmetry.  Given the problems encountered in relating 
the PG phase to precursor pairing fluctuations above $T_{c}$~\cite{[3]},  
a new broken symmetry state, with $d$-density wave ($d$-DW) order, was 
proposed~\cite{[4]}; the $d$-DW state was hypothesised to {\it compete} with
$d$-SC, being favoured in the underdoped region above $T_{c}$.
$(ii)$ On the other hand, careful measurements~\cite{[1]} reveal that the 
cuprates remain insulating (at very low $T$ if one suppresses $d$-SC by high 
magnetic fields) almost up to optimal doping. Given the amount of disorder 
involved, this insulator would be a Mott-Hubbard-Anderson insulator.  If 
the Mott transition occurs around $x_{opt}$, a natural question would be 
to inquire about its relation to the maximum of $T_{c}(x)$ at the same point.  
Does the Mott insulator then have $d$-wave order beginning from the 
undoped ($x=0$) state? (AF/SC long range order is precluded in 2D
by the Mermin-Wagner theorem, but $d$-wave order is not).
 How does this evolve to a nFL metallic state upon 
hole doping?

An attempt to reconcile both $(i)$ and $(ii)$ above would require that the 
Mott transition is an insulator-metal (IM) transition from a $d$-wave 
insulator to a $d$-wave metal. This gives rise to the question:  Is the 
quantum critical point (QCP) invoked to understand cuprates~\cite{[6]} the 
one associated with such a $T=0$ Mott transition?  (Recall that the $T-x$ 
phase diagram shows that AF-Neel order is destroyed well {\it before} 
$T_{c}(x)$ achieves its maximum around $x_{opt}$). If so, one would expect 
strongly divergent, $d$-wave charge/spin fluctuations around this transition 
(around $x_{opt}$) to naturally mediate high-$T_{c}$ SC of $d$-wave type. 
This would constitute a very new picture of the physical response of cuprates.

Additionally, strong dynamical spectral weight transfer (SWT), a ubiquitous 
feature of correlated systems, is exhibited clearly~\cite{[6]} as 
the Mott insulator is doped (this occurs over a region of $O(4.0~eV)$, 
and can only be caused by electronic correlations). Theoretically, non-local 
effects giving rise to $d$-wave correlations can be accessed only by 
cluster-DMFT (see review in~\cite{[7]}).  Such studies indeed find good
qualitative agreement with the $T-x$ phase diagram for the cuprates. However, 
to our best knowledge, the compelling link between the unusual nFL responses, 
the $T=0$ Mott transition around $x_{opt}$, and the high-$T_{c}$ 
SC of $d$-wave type still remains to be described consistently.

Here, with this motivation, we study the Mott transition in a two-dimensional 
model (without chemical disorder) for cuprates. We construct 
a new cellular-dynamical mean field approximation (C-DMFA) and use it to study 
the doping-driven Mott transition from a $d$-wave Mott insulator to a 
$d$-wave non-FL metal. We consider the extended Hubbard 
model (EHM) in the strong coupling limit as an effective model for Cu-O 
layers in cuprates~\cite{[8]}. The Hamiltonian is

\bn
\label{eq1}
H&=&-t\sum_{<i,j>,\sigma}(c_{i\sigma}^{\dag}c_{j\sigma}+h.c) 
+ U\sum_{i}n_{i\uparrow}n_{i\downarrow} \nonumber \\ 
&+& V\sum_{<i,j>}n_{i}n_{j} - E_F \sum_{i\sigma}n_{i\sigma} \;,
\en
where $n_{i}=\sum_{\sigma}n_{i\sigma}$, and $U>>V,t$, and $E_F$ is the 
chemical potential.  

Next, we formulate a new Dynamical Cluster Approximation, which we christen 
Cellular Dynamical Cluster Approximation (CDCA).  Given indications from 
several experiments~\cite{[2],[3]}, our aim is to treat inter-site 
electronic correlations 
{\it exactly} on a single plaquette (2x2) and consider the inter-plaquette 
correlations (resulting from carrier hopping and n.n interactions) within 
the CDCA.  This is done by mapping the original EHM to an electronic model 
describing coupled plaquettes, and is a multi-stage procedure as described 
below: 
(1) Consider the EHM on a single 2x2 square plaquette, $C$.  In 
the limit $U\rightarrow\infty$, the EHM reads

\bn
H &=& -t\sum_{\sigma, <i,j>=1}^{4}(X_{i\sigma}^{\dag}X_{j\sigma}+h.c)
+ V\sum_{<i,j>=1}^{4}n_{i}n_{j}  \nonumber \\  
&-& E_F \sum_{\sigma,i=1}^{4}n_{i\sigma} \;,
\en
where the no-double occupancy constraint resulting from the 
$U\rightarrow\infty$ limit is implied, and appears as a local constraint 
on $H$. Using the Gutzwiller approximation~\cite{[9]}, this becomes an 
unconstrained fermion model, but with $V=V(x),t=t(x)$. 
(2) Construct the four possible symmetry-adapted linear combinations of 
the site fermions to ``plaquette-centered'' fermions: 
$a_{\mu\sigma}=U_{i\mu}c_{i\sigma}$ with $i=[1,2,3,4]$ for 
a $4$-site square. Explicitly, 
$a_{1\sigma}=\frac{1}{2}(c_{1\sigma}+c_{2\sigma}+c_{3\sigma}+c_{4\sigma})$, 
with
$a_{2\sigma}=\frac{1}{2}(c_{1\sigma}-c_{2\sigma}+c_{3\sigma}-c_{4\sigma})$, 
and 
$a_{3\sigma}=\frac{1}{2}(c_{1\sigma}+c_{2\sigma}-c_{3\sigma}-c_{4\sigma})$, 
and finally, 
$a_{4\sigma}=\frac{1}{2}(c_{1\sigma}-c_{2\sigma}-c_{3\sigma}+c_{4\sigma})$.  
Notice that $a_{1}$ transforms as an $s$-wave operator, $a_{2}$ as a $d$-wave 
operator, while $a_{3},a_{4}$ transform like $p$-wave operators. The one 
plaquette Hamiltonian becomes ($\mu$ is centered on $C$)

\bn
\nonumber
H_{C} &=& -2t\sum_{\mu,\sigma}(n_{a1\mu\sigma}-n_{a2\mu\sigma}) 
+ V\sum_{\mu,\alpha,\beta}n_{a\mu\alpha}n_{a\mu\beta} \\
&-& E_F \sum_{\mu,\alpha}n_{a\mu\alpha} \;,
\en
where $\alpha,\beta=[1,2,3,4]$, and $n_{a\alpha}=a_{\alpha}^{\dag}a_{\alpha}$.
Thus, for half-filling, $t<0$, and $U\rightarrow\infty$, the exact spectrum 
of the EHM on an isolated square consists of four singly occupied plaquette 
states, $E_{1}=-2t, E_{2,3}=0, E_{4}=2t$, which we identify with the ``lower 
Hubbard band'' states, and the corresponding states raised upwards by $V$, 
which we identify with the ``upper Hubbard band'' (charge transfer) states. 
In analogy with the usual Hubbard model, the UHB states correspond to 
high-energy states describing the dynamics of an electron hopping onto a 
given local ``site'' (plaquette) in a specified ``orbital''  when the other 
``orbitals'' are already occupied.  $H_{C}$ in the plaquette basis is exactly 
similar to the local limit of a four-orbital Hubbard model with 
$U\rightarrow\infty$, and a fictitious ``crystal field'' term which raises 
the $d$-wave state above the doubly degenerate $p_{x,y}$ and extended-$s$ 
states. 
(3) We now connect $C$ to other plaquettes via the hopping, $t$, 
and n.n coupling, $V$, in the original EHM. To do this, we split the 2D 
lattice into plaquettes covering the whole lattice. The hopping and n.n 
coupling between neighboring sites on {\it different} neighboring plaquettes 
(in terms of original site fermions) are then expressed in terms of the 
plaquette fermions ($a_{\mu\sigma}$). In the 2D square geometry, $V,t$ 
acting {\it between} n.n plaquettes means that sites $(2,3)$ on plaquette 
$\mu$ are connected to sites $(1,4)$ on $(\mu+1)$ along $x$, while $(1,2)$ 
on $\mu$ are connected to $(3,4)$ on $(\mu+1)$ along $y$. Namely, the 
Hamiltonian coupling the given plaquette $\mu$ to the rest of the lattice is

\bn
H_{t}' &=& -t\sum_{\mu\sigma}(c_{1\mu\sigma}^{\dag}c_{2,\mu+e_{x},\sigma}
+c_{4\mu\sigma}^{\dag}c_{3,\mu+e_{x},\sigma}+h.c)  \nonumber \\ 
&-& t\sum_{\mu\sigma}(c_{1\mu\sigma}^{\dag}c_{4,\mu+e_{y},\sigma}
+c_{2\mu\sigma}^{\dag}c_{3,\mu+e_{y},\sigma}+h.c)  \;,
\en
and
\bn
H_{V}' &=& V\sum_{\mu}(n_{2,\mu}n_{1,\mu+e_{x}}+n_{3,\mu}n_{4,\mu+e_{x}}) 
\nonumber \\
&+& V\sum_{\mu}(n_{4,\mu}n_{1,\mu+e_{y}}+n_{2,\mu}n_{3,\mu+e_{y}}) \;. 
\en

In terms of the plaquette fermions, $H_{t}'+H_{V}'\equiv H'$ becomes
\bn
H' &=& -\sum_{\alpha\beta\sigma}t_{\alpha\beta}(a_{\mu\alpha\sigma}^{\dag}
a_{\mu+e_{x,y}\beta\sigma}+h.c) + \frac{V}{16}\sum_{\mu,\alpha\beta}
n_{\mu\alpha}n_{\mu+1,\beta}  \nonumber \\ 
&-&\frac{V}{8}\sum_{\mu\sigma\sigma'}(a_{1\mu\sigma}^{\dag}a_{2\mu\sigma'}^{\dag}a_{2,\mu+e_{x,y}\sigma'}a_{1,\mu+e_{x,y}\sigma} \nonumber \\
 &+& {1,2}\rightarrow {3,4}+h.c) + other\; terms
\en
where $\langle \mu,\nu\rangle$ now denote neighboring clusters. Here, 
$t_{\alpha,\alpha}=(t/2)(1,-1,1,-1)$ for $\alpha=[1,2,3,4]$, and $t_{13,24}=t/2(1,-1)$ along $x$ and $t/2(-1,1)$ along $y$. Our total 
Hamiltonian in terms of plaquette centered fermions is multi-orbital Hubbard 
model with the intra-orbital Hubbard $U\rightarrow\infty$, the 
``inter-orbital'', plaquette-local Coulomb interaction $V$, along with an 
explicit non-local, interaction term. This term connects neighboring 
{\it clusters}, i.e, it is of $O(1/D^{2})$. We decouple it in the 
generalised Hartree-Fock scheme (exact to $O(1/D^{2})$). The {\it other terms}
are unimportant near half-filling (they couple ex-$s$ and $p_{x,y}$ states), 
and are neglected henceforth.

Two $d$-wave order parameters are obtained: the particle-hole average, 
$\Delta_{d1}=\langle a_{1\mu\sigma}^{\dag}a_{2\mu\sigma}+a_{3\mu\sigma}^{\dag}a_{4\mu\sigma}\rangle$ and the $d$-wave pairing average, 
$\Delta_{12}=\langle a_{1\mu\sigma}^{\dag}a_{2\nu-\sigma}^{\dag}\rangle$, as 
required. A multi-orbital dynamical mean field approximation (MO-DMFA) 
for this Hamiltonian is exactly equivalent to a cluster-DFMA for the original 
EHM on the 2D square lattice. Thus, we solve $H=H_{C}+H'$ using the MO-DMFA. 

In order to make progress, we use insights gained from detailed analyses of 
the general multi-orbital problem in usual DMFA~\cite{[10],[11]}. Here, 
multi-orbital models without the inter-orbital, intersite pairing terms were 
solved using IPT and QMC, both leading to qualitatively similar conclusions. 
The Mott insulator has ferro-orbital (in our case $d$-wave) order: this is 
a $d$-wave Mott insulator. Away from $n=1$, the highest-lying ($d$-wave) 
band becomes metallic, while the lower-lying (here, one or more of 
ex-$s,p_{x,y}$) bands remain Mott insulating. For our model, this would 
describe a first order Mott transition from a $d$-wave insulator to a 
$d$-wave metal for any $x=(1-n)>0$ in absence of chemical (doping-induced) 
disorder.  As we show below, this $d$-wave metal is {\it not} a Fermi liquid.  

We solve our cluster (mapped 4-orbital) problem using multi-orbital 
iterated perturbation theory (MO-IPT). As discussed in recent~\cite{[12]} 
work, the MO-IPT turns out to be an accurate solver for multi-orbital 
problems at arbitrary fillings, $n \le 1$. We refer the reader to relevant 
references~\cite{[10],[12]} for methodical details. We focus on the 
Mott-Hubbard I-M transition as a function of hole doping. The role of the 
$d$-wave p-h (PG) and SC orders are described in an accompanying 
work~\cite{[19]}. Here, we focus on the region of the $T-x$ phase diagram 
for cuprates without $d$-wave PG/SC order; i.e, on the strange metal phase 
terminating at $T=0$ at $x_{opt}$.

\begin{figure}[t]
\includegraphics[width=\columnwidth]{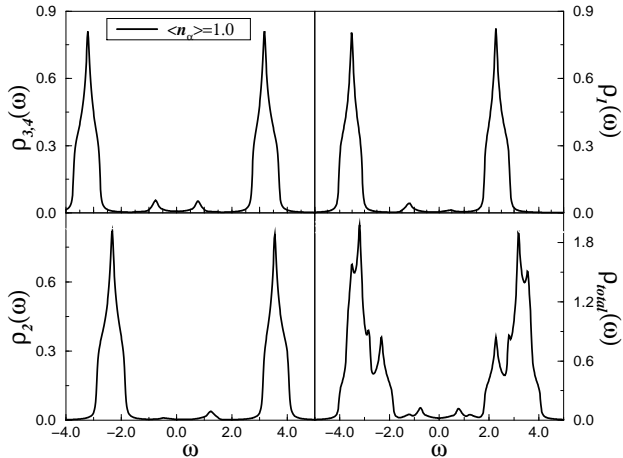}
\caption{The ``orbital resolved'' and the total many-particle density of 
states (DOS) for the $2D$ EHM at half-filling, $\langle n\rangle=1$.  Notice 
the appearance of ``orbital'' dependent Mott-Hubbard gaps, and the $d$-wave
insulator, as explained in the text.}
\label{fig1}
\end{figure}

\begin{figure}[t]
\includegraphics[width=\columnwidth]{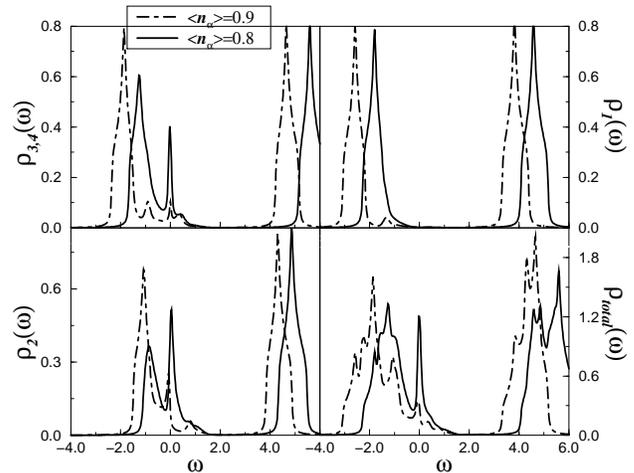}
\caption{Same as in Fig.(1), but for hole doping, $x=0.1$ (dashed lines) and
 $x=0.2$ (solid lines).  The ``orbital selective'' character of the Mott transition is explicitly seen, as is the fact that the ex-$s$-wave DOS remains Mott
localised in the metallic phase.}
\label{fig2}
\end{figure}

We now describe our results.  In Fig.~\ref{fig1}, we show the ``orbital'' 
resolved and total one-particle spectral function for our model with 
$V=4.0~eV, W=8t=4.0~eV$ for half-filling.  First, it is clear that the occupied states 
closest to $E_{F}(=0)$ have $d$-wave symetry (observe that 
$d \otimes s'=d, p_{x} \otimes p_{y}=d$). The Mott-Hubbard gap, 
$\Delta_{MH} \approx 1.0~eV$. Very interestingly, different ``orbital'' 
DOS have 
different Mott gaps: this is a benchmark feature of the ($d$-wave) ground 
state breaking the lattice point-group symmetry.  In terms of original 
$c$-fermions, these features result directly from an explicit (and exact 
to $O(1/D)$) consideration of the competition between strong, {\it intersite} 
correlations and one-electron hopping in the strong coupling ($U>>V,t$) 
limit of the EHM.  Together with the rich Hubbard band structures in 
Fig.~\ref{fig1}, it describes the internal structure of the Mott-Hubbard 
gap (inaccessible in $d=\infty$). We predict that polarised optics, XAS, 
Raman measurements would show anisotropic spectral features reflecting 
those found in the DOS: this may already have been seen.  

To extend our calculation to $n<1$, the MO-DMFT equations need to be 
constrained by the Friedel-Luttinger sum rule, 
$4n=4(1-x)=\int_{-\infty}^{E_{F}}\rho_{total}(\omega)d\omega$ for 
our cluster problem.  $E_{F}$ is then determined self-consistently within 
the MO-DMFT loop at each stage using $\rho_{total}(\omega)$ computed in each 
iteration, and the process is iterated to convergence.  The spectral 
functions for $n<1$ are shown in Fig.~\ref{fig2}. 
As expected generally, hole doping transfers dynamical spectral weight 
over large energy scales $O(4.0~eV)$ from the Hubbard bands into 
the Mott gap. The MIT is discontinuous, with the carrier density, 
$\langle n\rangle$, jumping discontinuously from zero to a finite value 
off half-filling (not shown), as generically expected within 
MO-IPT~\cite{[10]}. What is more interesting, and crucially important, 
is that this SWT is also strongly ``orbital'' dependent: clearly, our 
results show that the highest-lying $d$-wave DOS becomes the most metallic, 
the $p_{x,y}$ DOS much less so, while the ex-$s$-wave DOS remains Mott 
insulating.  In ``orbital'' language, this is an ``orbital selective'' 
Mott transition, discussed in detail in recent LDA+DMFT work on real 
oxides~\cite{[9],[10]}. In our CDCA applied to the 2D EHM, this feature 
directly follows from the strong, dynamical scattering {\it between} 
electronic channels with different ``orbital'' symmetries in a 
situation where these are split on the single cluster (reminiscent of 
crystal-field splitting in usual multi-orbital models~\cite{[10],[11]}).

A very remarkable fact now emerges.  Since the ex-$s$-DOS remains Mott 
insulating in the metal, strong scattering between the quasi-itinerant 
$d$-DOS and the localised ex-$s$-DOS results in an {\it exact} mapping of 
our problem to an effective Falicov-Kimball model~\cite{[10],[11]}, with 
resulting X-ray edge (XRE) behavior. The $d$-wave DOS then shows a power-law 
fall off at low energy in contrast to the smeared lorentzian of a FL metal: 
$\rho_{d}(\omega) \simeq |\omega|^{-(1-\alpha)}$, with 
$\alpha=tan^{-1}(V(x)/W(x))$. Hence, so does the total DOS. The corresponding 
self-energy is $\Sigma(\omega) \simeq |\omega|^{(1-\alpha)}$, with 
$\alpha<1$ for $V>0$~\cite{[13]}. This implies no FL quasiparticle residue: 
$Z_{FL}=0$, and the metallic state is a non-Fermi liquid. Amazingly, this 
is exactly in agreement with Anderson's proposal~\cite{[14]} for the 
breakdown of FL theory in the 2D Hubbard model. It is indeed remarkable 
that a close numerical estimation of the DOS near $E_{F}(=0)$ gives 
$\alpha\simeq 0.2,0.1$ for $x=0.1,0.2$. To our best knowledge, this is 
the first explicit demonstration of Anderson's conjecture within C-DMFT 
schemes. Recall that XRE behavior was suggested in the past~\cite{[15]} in 
multi-channel, single-impurity Anderson (and their $D=\infty$ lattice 
counterparts) models.

Armed with this insight, several unusual features of optimally doped cuprates 
are understood naturally.  The photoemission intensity, as measured in PES, 
measures the one-particle Green's function in the sudden approximation: hence, 
$I_{PES}(\omega) \simeq \omega^{-(1-\alpha)}$ as $\omega\rightarrow E_{F}$, 
as observed~\cite{[14]}.  The in-plane optical conductivity, 
$\sigma_{xx}(\omega)\simeq \omega^{-(1-2\alpha)}$ from a direct scaling 
argument, and drops like $\omega^{-0.6}(x=0.1),\omega^{-0.8}(x=0.2)$ very 
close to that measured experimentally~\cite{[16]}. Using an ``extended 
Drude'' fit~\cite{[16]}, we also obtain that the scattering rate, 
$\tau^{-1}(\omega) \simeq \omega^{(1-2\alpha)}$, as indeed observed. 
Further, the optical phase angle, 
$\phi=tan^{-1}(\sigma_{2}/\sigma_{1})=(1-2\alpha)\pi/2$, is 
independent of $\omega$, and increases with decreasing $x$, as seen.  
Finally, the electronic Raman scattering intensity in the $B_{1g}$ channel
follows from the local limit of the Shastry-Shraiman relation (valid in 
our MO-DMFT) for intraband transitions~\cite{[BS]}: 
$I_{R}^{el}(\omega)=\omega\sigma_{1}(\omega) \simeq \omega^{2\alpha}$, and 
is very weakly $\omega$-dependent, again as seemingly observed.  Very 
interestingly, this non-FL behavior extends up to rather high energies 
$O(1.0~eV)$, fully consistent with that used phenomenologically in optical 
analysis~\cite{[16]}.  Following Anderson~\cite{[18]}, the tunnelling 
spectrum will also exhibit power-law anomalies, reflecting that of the DOS.

What is the dominant instability of this singular non-FL phase?  Using the 
XRE mapping, it is clear that the ``excitonic''correlation function,
$\chi_{12}"(\omega)=\int_{-\infty}^{\infty}d\tau e^{i\omega\tau}\langle a_{1\sigma}^{\dag}a_{2\sigma}(\tau);a_{2\sigma'}^{\dag}a_{1\sigma'}(0)\rangle \simeq
|\omega|^{-(2\alpha-\alpha^{2})}$ is also singular.  This implies soft 
$d$-wave $p-h$ modes, and so the dominant instability will be to a $d$-wave 
ph pseudogapped state.  Details will be presented elsewhere~\cite{[19]}.

In conclusion, we have formulated a new cluster-DMFT to treat the effects 
of {\it intersite} electronic correlations (exactly) on the Mott transition 
in the 2D EHM. The $d$-wave Mott-Hubbard insulator found for $n=1$ 
undergoes a first-order Mott-Hubbard transition to a $d$-wave, non-FL metal 
upon hole doping. Remarkably, in absence of the $d$-pseudogap, this nFL metal 
shows low energy singular behavior of the one-particle DOS/$d$-wave 
``excitonic'' correlator, exactly as postulated by Anderson. Our results 
provide an appealing, microscopic explanation for a host of normal state 
anomalies characterising the ``strange metal'' phase of cuprates near 
optimal doping.  

M.S.L. thanks Prof. P. Fulde for advice and support at the MPIPKS, Dresden 
and EPSRC (UK) for financial support.  L.C.'s work was done under auspices 
of the SfB608 of the DFG.

\end{document}